\PassOptionsToPackage{numbers,square,sort&compress}{natbib}
\documentclass{article}

\usepackage[final]{neurips_2026}

\usepackage{graphicx}
\usepackage{hyperref}
\usepackage{booktabs}
\usepackage{amsmath}
\usepackage{amssymb}
\usepackage{xcolor}
\usepackage{subcaption}
\usepackage{comment}

\begin{document}

\title{
Towards Engineering Scaling Laws \\with Pretraining Data Composition
}

\author{
  Jan-Lucas Uslu \\
  Department of Applied Physics \\
  Stanford University \\
  Stanford, CA 94305, USA \\
  \texttt{uslu@stanford.edu}
  \And
  Kevin Greif \\
  Department of Physics and Astronomy \\
  University of California, Irvine \\
  Irvine, CA 92697, USA \\
  \texttt{kgreif@uci.edu}
  \And
  Daniel Whiteson \\
  Department of Physics and Astronomy \\
  University of California, Irvine \\
  Irvine, CA 92697, USA \\
  \texttt{daniel@uci.edu}
  \And
  Benjamin Nachman
  \\
  Department of Particle Physics and Astrophysics\\
  SLAC National Accelerator Laboratory \\
  Stanford University \\
  Menlo Park, CA 94025, USA \\
  \texttt{nachman@stanford.edu}
}

\maketitle

\begin{abstract}
    Neural scaling laws describe how model performance improves as a power law in compute, model size, and dataset size.
     While well-established for large language models, these relationships are  emerging for large models in particle physics.
     As with language, empirical studies show that the performance scales as a power law.
    However, unlike natural language or image domains, fundamental physics has high-fidelity simulators that produce synthetic data cheaply.
    This favors scaling regimes where additional data is cheaper than additional parameters, and allows the pretraining dataset itself to be engineered to influence the scaling.
     For the task of classifying hadronic jets produced in collisions of high-energy particle beams, we show that the scaling behavior can be engineered towards requiring more data rather than larger models by inclusion of pretraining data which is more diverse and better aligned with the downstream classification task. 
\end{abstract}

\section{Introduction}
\label{sec:introduction}

Neural scaling laws~\cite{kaplan2020scaling,hoffmann2022training} have emerged as a central tool for understanding and predicting the performance of large neural networks.
In the context of large language models (LLMs), these laws describe power-law relationships between loss and compute budget, model size, and dataset size, enabling compute-optimal training strategies.
For a fixed computing budget, for example, they predict whether more data or larger models will yield improved performance.
While these empirical laws demonstrate that more resources will result in better performance, they also show that returns diminish, making each additional gain increasingly expensive.

In the context of language or image learning, such studies focus on curation of existing datasets, as harvesting new datasets is expensive or even infeasible.
But in fundamental physics, new datasets can be generated via high-fidelity simulations which are already central to the scientific workflow, from experimental design to calibration and statistical analysis.
Beyond that, datasets can be generated explicitly for the purpose of machine learning and designed for specific learning objectives via judicious choice of the physical generative process.
This is a vital handle, as a growing number of foundation models~\cite{mikuni2024omnilearn, mikuni2025omnilearned, birk2024omnijet, Ho:2024qyf, li2024jetclass2, Golling:2024abg, Harris:2024sra, Katel:2024ygn, Bardhan:2025icr, Wildridge:2024yeg, hallin2025foundation,park2025fm4npp,Mikuni:2024qsr,Parker:2025kne,xia2025multimodalfoundationmodelcosmological,Hsu:2026sww,Young:2025qah,Young:2025qbx} and scaling law studies~\cite{batson2023scaling,vigl2026neural,ATL-SOFT-PUB-2026-002,bahl2026scaling} in particle, nuclear, and astrophysics show that large models are promising but can be computationally expensive.
Can the unique data-generation capability of fundamental physics be leveraged to engineer the scaling behavior of large models?

It has been already demonstrated that beyond simply scaling dataset and model sizes, the composition of the pretraining data can influence the scaling behavior~\cite{sorscher2022beyond}.
Datasets with greater {\it diversity}, a wider variety of examples, or better {\it alignment}, examples whose features transfer to the downstream task, can build richer feature representations during pretraining, shifting the compute-optimal allocation during fine-tuning towards a preference for more data rather than larger models.
Can the composition of the pretraining data be engineered to shift compute-optimal scaling towards data-favoring regimes?

In this paper, we explore this question in the context of hadronic jets at particle colliders.
Jets are streams of fast-moving particles resulting from the radiation pattern of quarks and gluons produced from reactions that exchange a large amount of energy.
These objects are ubiquitous at high-energy colliders and have served as a baseline for machine learning method development due to their prevalence, relevance, and complexity.
We use the well-studied JetClass-II dataset~\cite{li2024jetclass2} for pretraining using a multi-class supervised objective~\cite{mikuni2024omnilearn} and  fine-tune on the original JetClass dataset~\cite{Qu2022mxj}.
To explore the impact of the pretraining composition on the scaling behavior, we use various subsets of the JetClass-II for pretraining.
In particular, we vary the diversity and alignment of the pretraining data by including more or less examples from jets generated with particles beyond the Standard Model, which are both more complex than typical jets produced from a single quark or gluon and better aligned with the downstream task of jet classification.
This is only one of many possible notions of diversity and alignment and we only study one pretraining setup and one fine-tuning task.
Future studies can expand on this initial exploration in many directions.

This paper is organized as follows.
Section~\ref{sec:related} briefly describes related work in fundamental physics as well as in the general machine learning literature.
The technical aspects of our study are documented in Sec.~\ref{sec:methods}.
Numerical results are presented in Sec.~\ref{sec:results} followed by some discussion in Sec.~\ref{sec:discussion}.
The paper ends with conclusions and outlook in Sec.~\ref{sec:conclusions}.

\section{Related Work}
\label{sec:related}

\paragraph{Neural scaling laws.}
Power-law relationships between loss and scale were first characterized empirically across multiple domains by~\cite{hestness2017deep} with respect to dataset size.
Kaplan et al.~\cite{kaplan2020scaling} formalized these for language models, and the Chinchilla analysis in~\cite{hoffmann2022training} derived compute-optimal scaling strategies balancing model size and training tokens.
Theoretical grounding was provided in Ref.~\cite{sharma2020neural}, which linked scaling exponents to the intrinsic dimension of the data manifold, and Ref.~\cite{bahri2024explaining} identified distinct variance-limited and resolution-limited regimes.

\paragraph{Scaling laws in HEP.}
Ref.~\cite{batson2023scaling} provided an early systematic study of scaling laws in fundamental physics using jet classification, showing that six classifiers follow power-law scaling with dataset size on top quark-initiated versus light-quark/gluon initiated jet discrimination with exponents between 0.037 and 0.105.
Most directly comparable to our work, Ref.~\cite{vigl2026neural} performed a full Chinchilla-style analysis on JetClass using a Set Transformer, fitting the parametric loss model
\begin{equation}
    L(N, D) = L_\infty + \frac{A}{N^\alpha} + \frac{B}{D^\beta},
    \label{eq:powerlaw}
\end{equation}
where $L_\infty$ is the irreducible loss, $A/N^\alpha$ captures the model-capacity bottleneck, and $B/D^\beta$ captures the data bottleneck, obtaining $\alpha=0.44$ and $\beta=0.22$.
They found that the irreducible loss depends on input representation and that multi-epoch training modifies scaling~\cite{muennighoff2023scaling}.
The ATLAS collaboration performed a similar analysis on a larger and more realistic jet classification dataset and extracted $\alpha = 0.68$ and $\beta = 0.07$~\cite{ATL-SOFT-PUB-2026-002}.
Ref.~\cite{bahl2026scaling} connected scaling exponents for amplitude surrogates to the number of external particles.
Most recently, ref.~\cite{Amram:2026zzv} explored scaling laws for generative tasks in HEP.

Our work differs from these in that we specifically study how pretraining data composition, both diversity and downstream alignment, shifts scaling exponents, rather than analyzing scratch training alone.

\paragraph{Transfer learning and data composition scaling.}
In the machine learning literature, Ref.~\cite{hernandez2021scaling} showed that pretraining effectively multiplies the fine-tuning dataset, with the multiplier following a power law in model size whose exponents depend on the proximity of pretraining and fine-tuning distributions.
Reference~\cite{isik2025scaling} found that pretraining-downstream alignment determines whether scaling is beneficial.
Motivating our work, Ref.~\cite{cherti2023reproducible} demonstrated that training distribution changes modify scaling exponents even for identical architectures and 
Ref.~\cite{sorscher2022beyond} showed that data pruning can go beyond power-law scaling.

Transfer learning has been studied for a number of years in fundamental physics, with early studies using natural language pretraining~\cite{Chappell:2022yxd}.
One factor that may have contributed to the improved performance of the OmniLearned foundation model~\cite{mikuni2025omnilearned} over the OmniLearn foundation model~\cite{mikuni2024omnilearn} is the enhanced diversity in the pretraining dataset of the former model.
This paper presents the first systematic study, to our knowledge, of how pretraining composition affects scaling laws in fundamental physics.
\section{Methods}
\label{sec:methods}

\subsection{Model Architecture}
\label{sec:model}

For all experiments, we use a generic transformer architecture which processes jet data as a point-cloud.
Each jet is represented as a set of constituent particles, with each particle characterized by 4 kinematic features, 4 trajectory displacement features, and 1 particle identification feature as in Ref.~\cite{Qu2022mxj}.
The model employs a learnable class token, pre-norm transformer blocks with SwiGLU~\cite{shazeer2020glu} feed-forward networks and does not use positional encodings, as the input is a set rather than a sequence.
To span a range of model sizes, we fix the depth at 4 layers and the attention head dimension at 8 while varying the embedding dimension from 8 to 512 and the number of heads, yielding 12 model sizes from approximately 3K to 10.5M parameters, spanning over three orders of magnitude in parameter count.

\subsection{Datasets}
\label{sec:datasets}

\paragraph{Pretraining: JetClass-II.}
We use the JetClass-II dataset~\cite{li2024jetclass2} for pretraining, which contains 188 classes with simulated jets initiated by a wide array of Standard Model (SM) and Beyond the Standard Model (BSM) processes.
These classes are grouped into three large buckets: jets initiated by light quarks or gluons (QCD), jets initiated by the two-body decay of a BSM resonance (2-prong, labeled {\it res2p}), and jets initiated by the three- or four-body decay of BSM resonances (3/4-prong, labeled {\it res34p}).
The 2-prong and 3/4-prong buckets contain only BSM resonance decays; SM resonance decays ($W$, $Z$, top, Higgs) are not part of JetClass-II.

The class labels for the QCD bucket are defined using the number of bottom ($b$), charm ($c$) and strange ($s$) quarks in the truth parton list prior to hadronization, while the class labels for the BSM buckets are defined using the first generation decay products of BSM resonances.

We define four pretraining subsets:
\begin{itemize}
    \item \textbf{QCD}: QCD jets only, comprising 17 classes
    \item \textbf{QCD + res2p}: QCD plus 2-prong resonance decays, with an approximate 40\%/60\% split between QCD and res2p
    \item \textbf{QCD + res34p}: QCD plus 3/4-prong resonance decays, with an approximate 20\%/80\% split between QCD and res34p
    \item \textbf{QCD + res2p + res34p}: QCD plus all resonant decays, with an approximate 15\%/20\%/65\% split between QCD, res2p, and res34p
\end{itemize}
The split ratios are selected based on the original distribution of jets across the three buckets in JetClass-II.
BSM decays populate regions of phase space that QCD jets do not, broadening the support of the pretraining distribution.

Throughout this paper, we use {\it diversity} to refer to the variety of physics processes in the pretraining corpus, measured loosely by the number of distinct process classes and the range of kinematic configurations they span.
We use {\it alignment} to refer to the overlap between pretraining and fine-tuning data in features relevant to the downstream task: prong multiplicity, mass scale, and substructure.
The BSM-augmented subsets are both more diverse and better aligned than QCD-only; the present experiments do not separate these effects.

\paragraph{Downstream: JetClass.}
The downstream task is 10-class jet classification using the JetClass Dataset~\cite{Qu2022mxj}.
In contrast to the parton-flavor jet classes of JetClass-II, the classes of JetClass are only based on the initiating particle.
These are light quarks or gluons, top quarks, $W/Z$ bosons, or Higgs particles.

The ten JetClass classes correspond to four distinct prong topologies: light quarks and gluons produce 1-prong jets; $W/Z\rightarrow qq$ and $H\rightarrow bb/cc/gg$ are 2-prong; $t \rightarrow bqq$ is 3-prong; and $H \rightarrow WW\rightarrow qqqq$  is 4-prong.
The downstream task is therefore largely one of identifying prong multiplicity, estimating mass scales, and tagging heavy flavor.
QCD-only pretraining is therefore aligned with the $q/g$ classes  but provides little signal for the multi-prong resonance classes that dominate the downstream task.
BSM-augmented pretraining adds coverage of the multi-prong resonance classes that QCD-only misses.

We use the same task and evaluation protocol as Ref.~\cite{vigl2026neural} to ensure comparability, which involves single-epoch fine-tuning on the full JetClass training set without data repetition.

\subsection{Training Protocol}
\label{sec:protocol}

\paragraph{Pretraining.}
Models are pretrained on each subset for 200{,}000 iterations using AdamW~\cite{loshchilov2019decoupledweightdecayregularization} with a learning rate of $10^{-3}$ and batch size 128, ensuring identical total token~\footnote{The input particles are the tokens - they are not tokenized.} counts across all subsets so that only the composition differs.
The number of iterations and batch size together imply that each model is pretrained on roughly 25.6 million jets.
Warmup is applied for the first 1{,}000 iterations.

\paragraph{Fine-tuning.}
When fine-tuning on the downstream task, we transfer the pretrained backbone weights and reinitialize the classification head.
Models are fine-tuned for 600{,}000 iterations on JetClass without data repetition, following the single-epoch protocol of Ref.~\cite{vigl2026neural}.
We use warmup for the first 5{,}000 iterations of fine-tuning, but otherwise keep all hyperparameters identical to pretraining.
Scratch baselines (no pretraining) follow the same single-epoch protocol.

\begin{table}[!ht]
    \centering
    \caption{Compute-optimal scaling exponents extracted from power law fits to $N^*$ and $D^*$ as a function of compute.
    Uncertainties from bootstrap resampling (100 samples, 80\% subsample fraction) of the test dataset, from which the test loss is recalculated.
    The scaling exponents are extracted from each bootstrap and the standard deviation across the ensemble is taken as the uncertainty.}
    \label{tab:exponents_approach1}
    \renewcommand{\arraystretch}{1.3}
    \begin{tabular}{lcc}
        \toprule
        Pretraining subset & $a$ ($N^*$) & $b$ ($D^*$) \\
        \midrule
        Scratch              & $0.517 \pm 0.002$ & $0.483 \pm 0.002$ \\
        QCD                  & $0.526 \pm 0.004$ & $0.474 \pm 0.004$ \\
        QCD + res2p          & $0.265 \pm 0.005$ & $0.735 \pm 0.005$ \\
        QCD + res34p         & $0.283 \pm 0.005$ & $0.717 \pm 0.005$ \\
        QCD + res2p + res34p & $0.224 \pm 0.005$ & $0.776 \pm 0.005$ \\
        \bottomrule
    \end{tabular}
\end{table}

\subsection{Scaling Law Analysis}
\label{sec:scaling}

Following Ref.~\cite{kaplan2020scaling}, we estimate compute as $C = 6  N  n_p  B  S$, where $N$ is the number of model parameters, $n_p = 40$ is the average number of constituents per jet, $B = 128$ is the batch size, and $S$ is the number of training iterations.
To extract the scaling exponents, we follow Approach 2 from Ref.~\cite{hoffmann2022training} and fit parabolas in the log-log space of test loss versus model size ($N$) and dataset size ($D$).
The minimum of a given parabola gives the compute-optimal model size $N^*$ or dataset size ($D^*$).
The compute-optimal scaling exponents $a$ and $b$ are then extracted from power-law fits to $N^*$ and $D^*$ as a function of compute:
\begin{align}
    N^* & \propto C^a, \\
    D^* & \propto C^b,
\end{align}
with the consistency condition $a + b \approx 1$.
We fit these power laws with least-squares regression in log-log space.

We additionally implemented Approach 3 from Ref.~\cite{hoffmann2022training}, which fits a parametric function to model the test loss as a function of $(N, D)$, as a secondary validation of the extracted compute-optimal scaling exponents.
More details are provided in the Appendix.

Following Ref.~\cite{hoffmann2022training} and Ref.~\cite{vigl2026neural}, we use cross-entropy loss over the test set as the loss used to extract the compute-optimal scaling exponents for both approaches.

\begin{figure}[t]
    \centering
    \begin{subfigure}[b]{\textwidth}
        \centering
        \includegraphics[width=\textwidth]{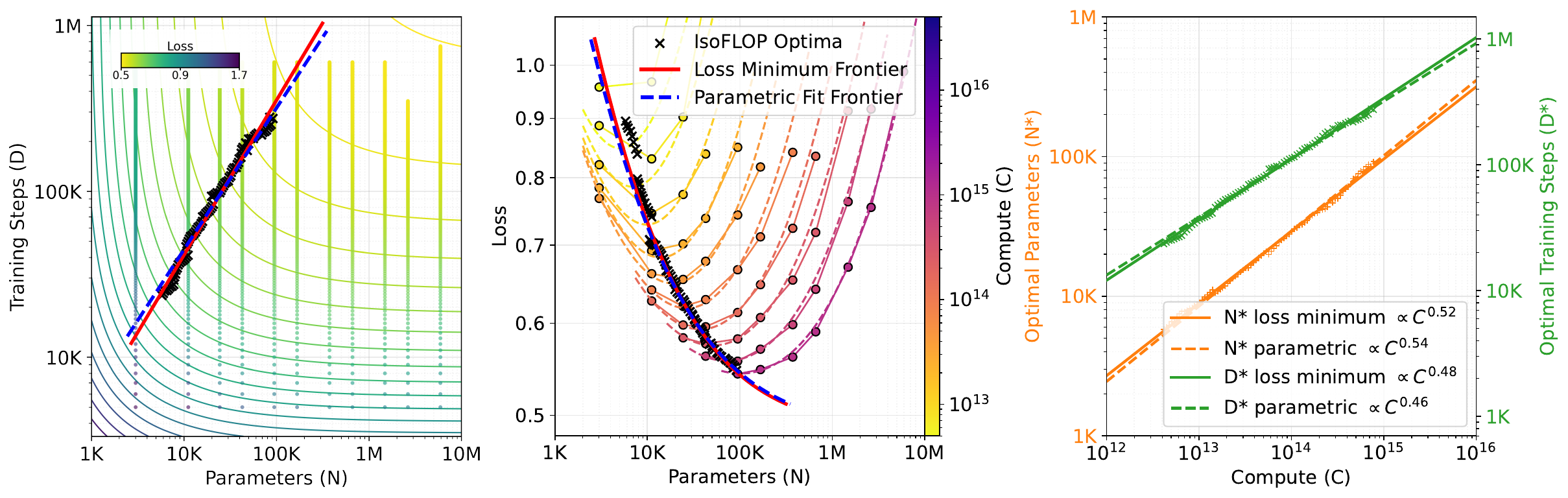}
        \caption{Scratch training.}
        \label{fig:scaling_scratch}
    \end{subfigure}\\[6pt]
    \begin{subfigure}[b]{\textwidth}
        \centering
        \includegraphics[width=\textwidth]{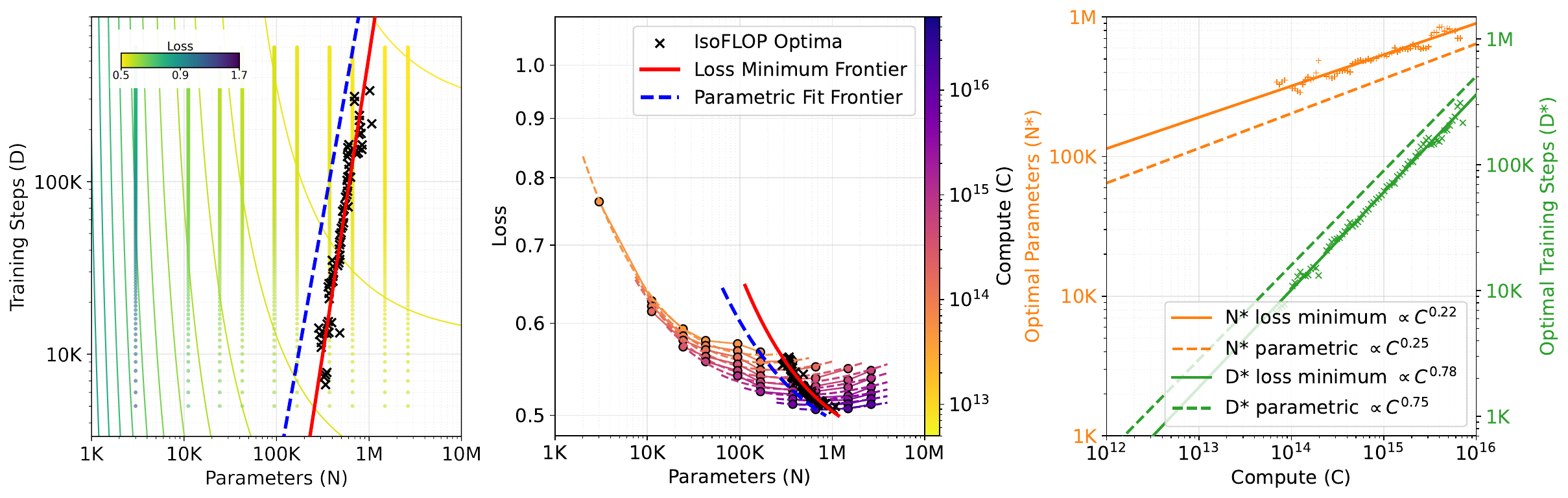}
        \caption{QCD + res2p + res34p pretraining.}
        \label{fig:scaling_bsm}
    \end{subfigure}
    \caption{Scaling analysis using test loss on JetClass after scratch training (a) and QCD + res2p + res34p pretraining using JetClass-II (b). \textbf{Left:} Loss contours $L(N,D)$ with the compute-optimal frontier from the parametric fit (Approach~3, blue dashed line) and the extracted IsoFLOP optima (Approach~2, crosses).
    The compute-optimal scaling tilts from a balanced $N$--$D$ allocation (a) to a strongly data-favoring regime (b). \textbf{Center:} IsoFLOP slices comparing the parametric fit (dashed) to the observed loss (solid), confirming good agreement across compute budgets. \textbf{Right:} Power-law fits to the optimal model size $N^*$ and dataset size $D^*$ as a function of compute.}
    \label{fig:scaling}
\end{figure}

\begin{figure}[p]
    \centering
    \begin{subfigure}[b]{\textwidth}
        \centering
        \includegraphics[width=\textwidth]{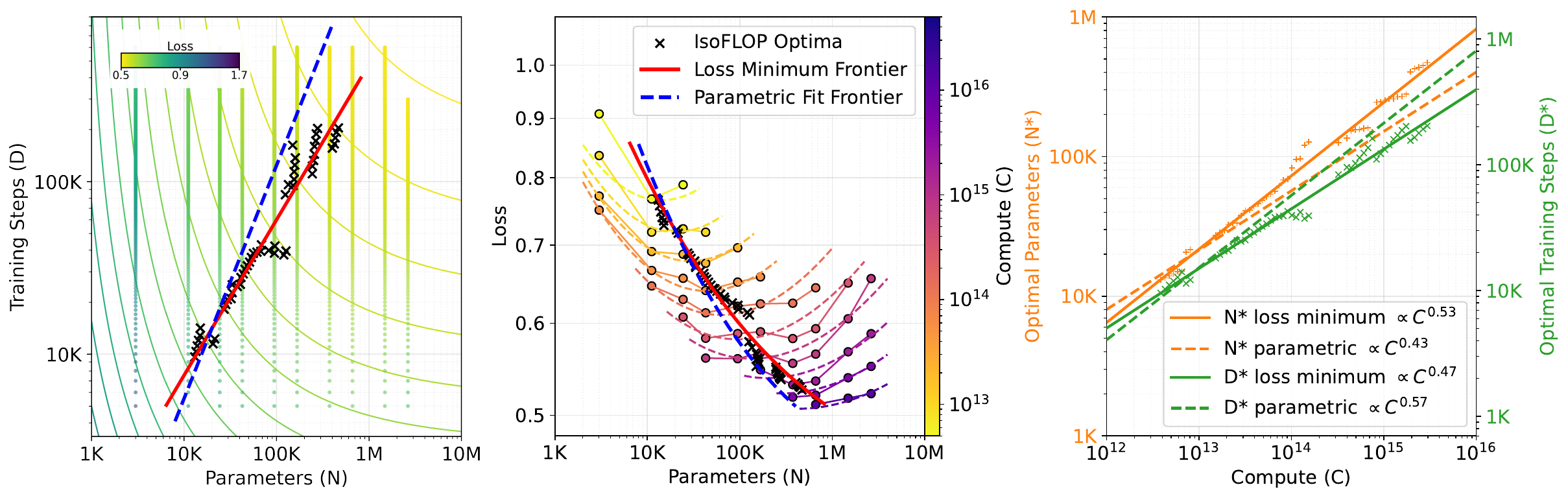}
        \caption{QCD-only pretraining.}
        \label{fig:scaling_qcd}
    \end{subfigure}\\[6pt]
    \begin{subfigure}[b]{\textwidth}
        \centering
        \includegraphics[width=\textwidth]{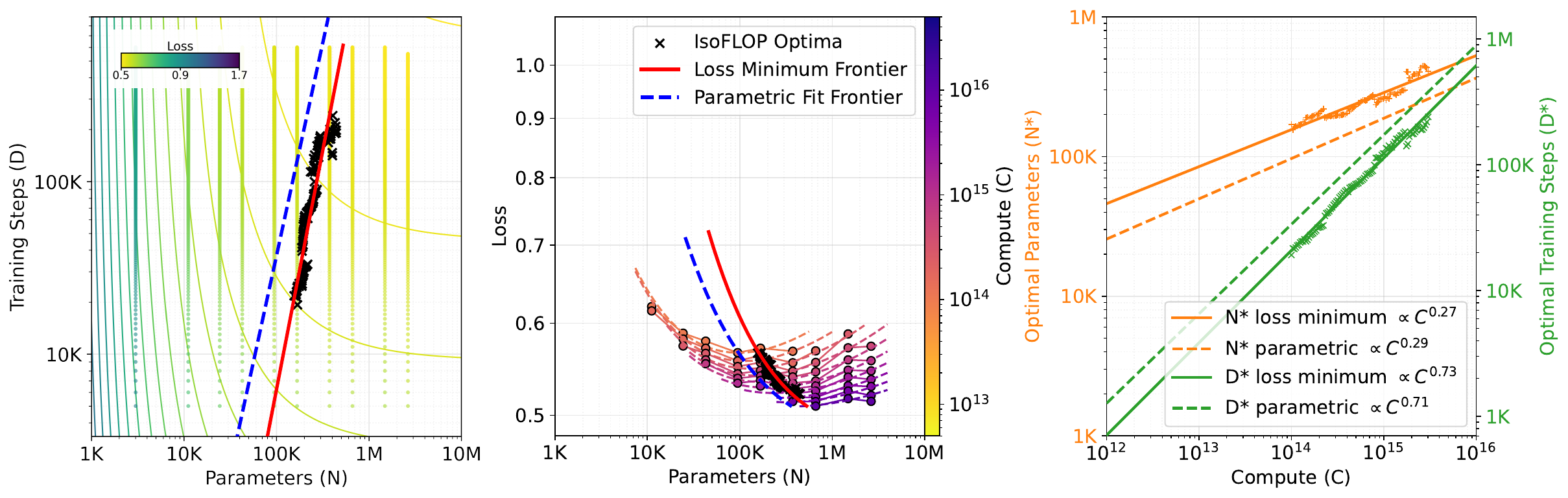}
        \caption{QCD + 2-prong resonance pretraining.}
        \label{fig:scaling_res2p}
    \end{subfigure}\\[6pt]
    \begin{subfigure}[b]{\textwidth}
        \centering
        \includegraphics[width=\textwidth]{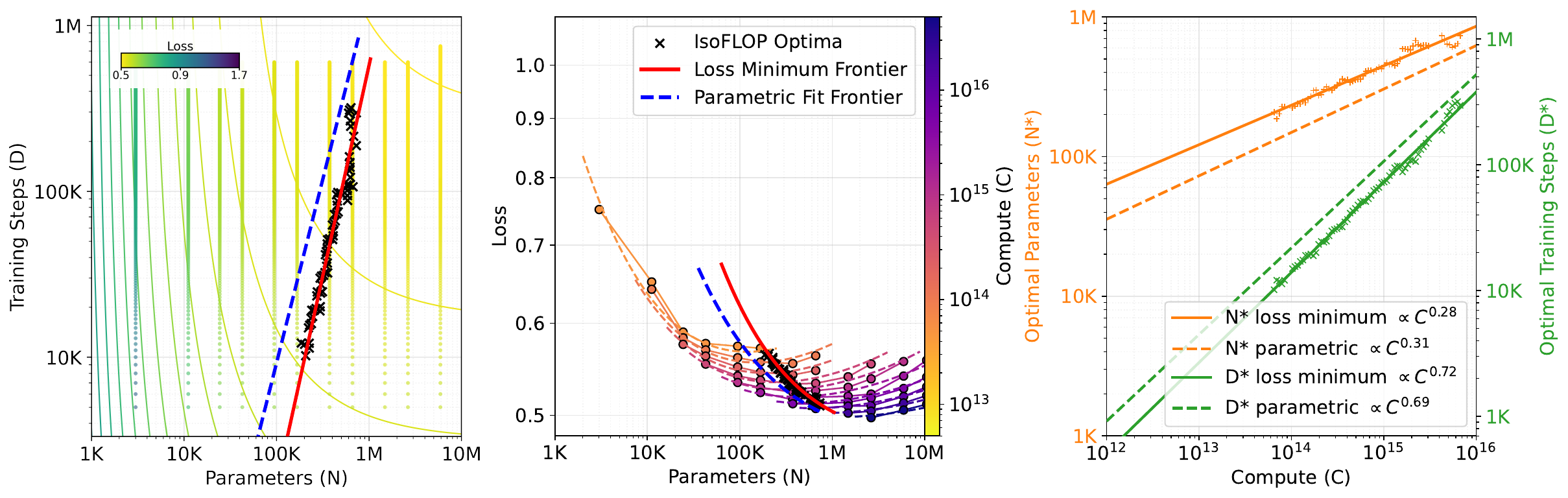}
        \caption{QCD + 3/4-prong resonance pretraining.}
        \label{fig:scaling_res34p}
    \end{subfigure}
    \caption{Scaling diagnostics for the intermediate pretraining configurations.
    The panels follow the same format as Figure~\ref{fig:scaling}: loss contours and compute-optimal frontiers (left), IsoFLOP slices (center), and power-law fits for the optimal model size and dataset size as a function of compute (right).}
    \label{fig:scaling_intermediate}
\end{figure}

\section{Results}
\label{sec:results}

Each (subset, embedding dimension) fine-tuning configuration is trained across 5 independent runs with different random seeds, for a total of 300 fine-tuning runs across the five configurations (scratch and the four pretraining subsets) and twelve model sizes.
Loss curves are smoothed, binned into windows of 1{,}000 iterations, averaged across runs, and monotonized via a running minimum following Ref.~\cite{hoffmann2022training}.
The first 5{,}000 iterations are discarded to exclude warmup transients.

\subsection{Scaling Laws without Pretraining}
\label{sec:scratch}

Training from scratch on JetClass, the scaling law analysis yields compute-optimal scaling exponents $a = 0.517 \pm 0.002$ and $b = 0.483 \pm 0.002$, reproducing the approximately equal allocation between model size and data found by Ref.~\cite{hoffmann2022training} who found $a = b = 0.5$.
Poor fit quality was observed for the parametric fit approach documented in the Appendix.
We suspect this fit quality would improve if more training runs were gathered, especially for models with larger numbers of parameters.
Despite this, similar compute-optimal scaling exponents of $a = 0.539 \pm 0.001$ and $b = 0.461 \pm 0.001$ were extracted from the secondary approach.

Figure~\ref{fig:scaling_scratch} shows the scaling analysis for scratch training.
The left panel shows the minimum test loss as a function of $(N, D)$, with compute-optimal training configurations marked with black crosses.
The power law fit to the data used to extract the compute-optimal scaling exponents is shown in red.
A very similar power law, obtained from the parametric fit approach documented in the Appendix, is shown in blue.
The center panel shows the observed test loss along iso-compute slices as a function of model size (solid) compared with the parametric-fit predictions (dashed).
Similar results were obtained for the analogous fits to test loss versus dataset size.
The right panel shows the power-law fits for $N^*$ and $D^*$ as a function of compute for both the IsoFLOP profile approach (solid) and the parametric fit approach (dashed).

\begin{figure}[t]
    \centering
    \includegraphics[width=\textwidth]{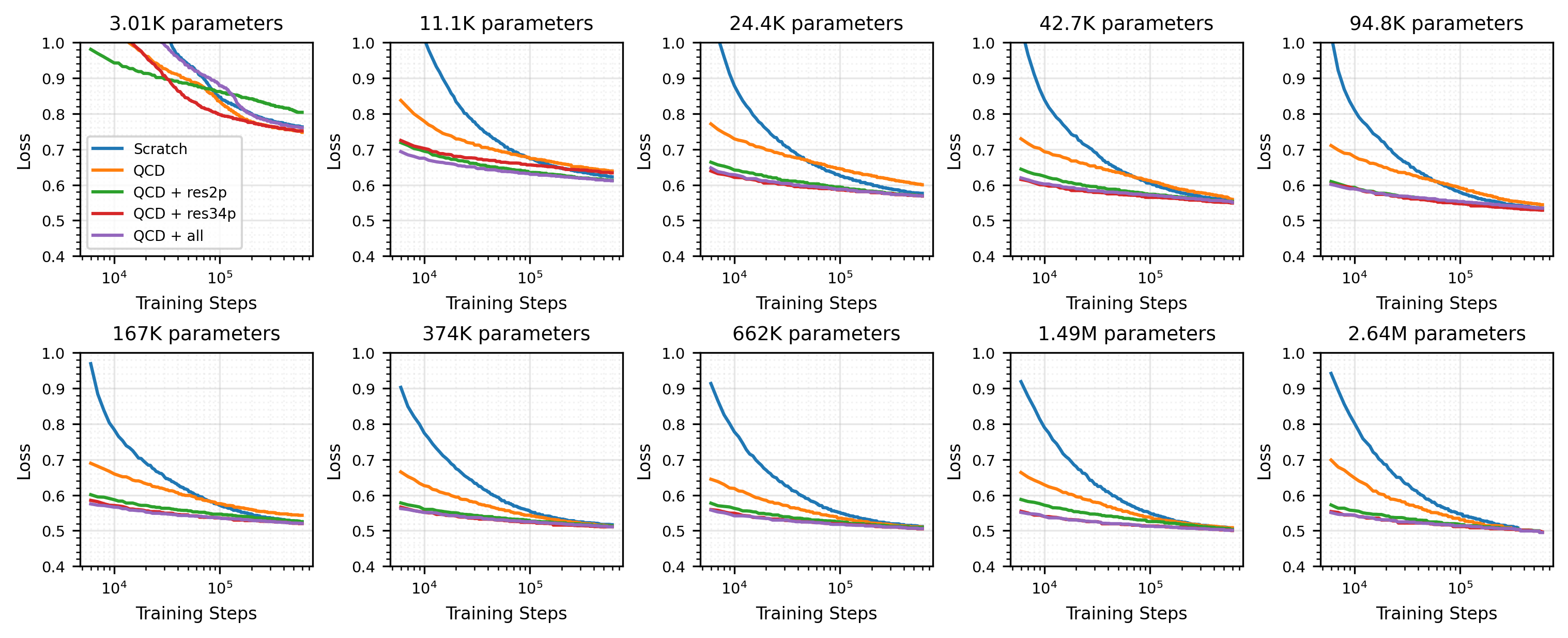}
    \caption{Fine-tuning loss curves as a function of training steps for each model size, comparing scratch training and all four pretraining configurations.
    BSM-enhanced pretraining consistently lowers the loss across all model sizes, with the benefit increasing for larger models.}
    \label{fig:loss_comparison}
\end{figure}

\subsection{Effect of Pretraining}
\label{sec:bsm_effect}

Comparing Figures~\ref{fig:scaling_scratch} and~\ref{fig:scaling_bsm} illustrates the impact of pretraining on compute-optimal scaling.
Pretraining on BSM-augmented JetClass-II subsets substantially alters the test loss values as a function of $(N, D)$, and produces a compute-optimal scaling where the FLOPs allocated to dataset size are increased much faster than the FLOPs allocated to model capacity.
Additional pretraining configurations are shown in Figure~\ref{fig:scaling_intermediate}, and show results in between the scratch training and full BSM-enhanced results shown in Figure~\ref{fig:scaling}.
Table~\ref{tab:exponents_approach1} summarizes the scaling exponents across all pretraining configurations.
Pretraining only on the QCD portion of JetClass-II produces only a marginal shift relative to scratch ($a = 0.526$ vs. $0.517$).
Adding 2-prong resonances (res2p) causes a large shift to $a = 0.265$, and including all resonances results in $a = 0.224$, a factor-of-2.3 reduction from the scratch training baseline.
The progression of pretraining datasets QCD to QCD+res2p to QCD+res2p+res34p tracks both increasing diversity and increasing topological overlap with the JetClass label space; we discuss the implications for disentangling these effects in Section~\ref{sec:discussion}.

The parametric fit approach to extracting compute-optimal scaling yields consistent results (see Table~\ref{tab:exponents_approach2}), confirming the shift is robust and not a fitting artifact.
However the pretraining configurations with BSM portions of JetClass-II show uniform offsets in the compute-optimal scalings extracted by both approaches, with the parametric fit approach yielding larger optimal dataset sizes and smaller optimal model sizes, as can be seen in the right panels of Figures~\ref{fig:scaling_bsm}, \ref{fig:scaling_res2p}, and \ref{fig:scaling_res34p}.
We expect these differences to result from the poor fit quality observed in the parametric fit approach.
Figure~\ref{fig:loss_comparison} shows the fine-tuning loss curves across all model sizes, confirming that BSM-enhanced pretraining consistently lowers the loss, with the benefit increasing for larger models.

In summary, pretraining on QCD jets alone reduces the loss but leaves the compute-optimal scaling essentially unchanged.
Adding jets produced by hadronic decays of BSM resonances shifts the compute-optimal scaling exponent for dataset size from $b \approx 0.48$ to $b \approx 0.78$, meaning more of the available compute budget should be allocated to training with additional data.

\section{Discussion}
\label{sec:discussion}

The key finding of this study is that in addition to lowering the test loss, pretraining alters the compute-optimal scaling exponents.
The shift from $b \approx 0.48$ (scratch) to $b \approx 0.78$ (QCD + res2p + res34p) implies that models pretrained on a corpus that is both more diverse and better aligned with the downstream task favor additional data above additional parameters at a fixed compute budget.
Recall that the BSM-augmented subsets differ from QCD-only along both axes:  jets produced by more process classes with broader phase-space coverage (diversity), as well as more overlap with the features of the jets used in the downstream task (alignment).
QCD-only pretraining produced only a marginal shift in the compute-optimal scaling exponents, showing that pretraining corpus composition affects the scaling much more strongly than pretraining alone.

The shift toward data-favoring compute-optimal scaling when pretraining on BSM-enhanced subsets can be interpreted as the pretrained representations substituting for model parameters, so that at a fixed compute budget a given downstream loss is reached with a smaller model~\cite{hernandez2021scaling}.
Compared with models trained on a naively constructed corpus, models pretrained on more diverse and aligned datasets reach a given loss with fewer parameters, so their model size does not need to scale as quickly with the compute budget.
Instead, additional compute resources should be used to train on more data assuming it is available.
In realistic jet classification tasks in particle physics, training data typically consists of synthetic data produced by simulators~\cite{ATLAS:2025dkv}.
This allows a high level of control over the construction of a pretraining corpus, ignoring the challenges presented by the computational expense of running the simulator itself.
We expect this simulation cost to be small compared to the cost of training large models, so that the data-favoring regime remains advantageous even when data generation is accounted for; a more involved training strategy could fold the simulation cost explicitly into the compute budget.
Our results suggest that this unique feature of particle physics datasets should be actively exploited when building foundation models.
In particular, foundation models pretrained on diverse and aligned datasets can be smaller than their naively trained counterparts, and the saved compute can be allocated to the generation of additional data by simulators for optimal scaling.

\section{Conclusions and Outlook}
\label{sec:conclusions}

We have shown that pretraining data composition significantly reshapes compute-optimal scaling laws for jet classification.
While pretraining on only light quark or gluon initiated jets approximately reproduces the exponents of scratch training ($a \approx b\approx 0.5$), including jets produced by BSM resonances in pretraining progressively shifts the compute-optimal scaling to favor additional data over larger models ($a \approx 0.22$, $b \approx 0.78$).
The practical implication is that when foundation models are pretrained on well-composed corpora, downstream fine-tuning compute is best spent on more data rather than larger models, a regime well-suited to fundamental physics, where simulated data is cheap relative to compute for large models.

Furthermore, pretraining data should be composed with attention to both diversity and alignment with anticipated downstream tasks, though additional work is required to verify these results generalize across different fine-tuning tasks as well as hold for larger dataset and model sizes.
More broadly, this work suggests that the physics inputs to foundation-model training, not just architecture and scale, are a meaningful design space for scientific machine learning, with pretraining composition engineering as one underexplored lever.

Future work should explore this direction, in addition to varying the pretraining compute budget to test whether larger-scale pretraining amplifies the data-favoring shift.
It would also be interesting to investigate alternative methods for creating dataset diversity from the Monte Carlo simulation and it may even be possible to co-optimize this with the pretraining objective.
A complementary direction is to define explicit, quantitative metrics for pretraining-data diversity and downstream alignment---for example, the sliced Wasserstein distance between datasets in a space of physically motivated observables such as $N$-subjettiness ratios or energy-flow polynomials---and to study scaling directly against these metrics rather than only against dataset size.
Associating dedicated scaling coefficients with diversity and alignment would turn the qualitative trends observed here into quantitative handles, and could frame the choice of which simulations to generate as an explicit exploration--exploitation problem.
This work opens up a new direction for maximizing the discovery potential of scientific foundation models by optimizing the physics inputs to build a richer representation of the data.

\section*{Code Availability}

The code used to produce the results in this paper, including the model implementation, training scripts, dataset preparation utilities, and the scaling-law analysis, is publicly available at \url{https://github.com/Jaluus/BSMScaling}~\cite{uslu2026bsmscaling_code}.

\section*{Acknowledgments}

We thank Steve Mrenna, Stephan Hoeche, Manuel Szewc, and Vinicius Mikuni for useful discussions early on in the formation of this project.
We are especially grateful to Manuel Szewc for detailed feedback on the manuscript.
BN is supported by the U.S. Department of Energy (DOE), Office of Science under contract DE-AC02-76SF00515.
KG and DW are supported by the DOE Office of Science.
This research used resources of the National Energy Research Scientific Computing Center, a DOE Office of Science User Facility supported by the Office of Science of the U.S. Department of Energy under Contract No. DE-AC02-05CH11231 using NERSC awards HEP-ERCAP0035546.

\section*{Appendix}
\label{sec:appendix}

For a complementary extraction method to the IsoFLOP profile approach described in the main text, we follow Ref.~\cite{hoffmann2022training} and fit the parametric loss model of Equation~\ref{eq:powerlaw} to the full $(N, D, L)$ dataset.
The compute-optimal scaling exponents are then derived analytically as $a = \beta/(\alpha + \beta)$ and $b = \alpha/(\alpha + \beta)$.
This approach yielded poor quality fits to the data, which is why the alternative approach is used for the main results.
Despite this the extracted compute-optimal exponents, shown for each pretraining configuration in Table~\ref{tab:exponents_approach2}, are very similar to those produced by the main approach.
This shows that our results are robust against methods used to extract scaling exponents.

The poor fit quality could result either from statistical noise in the minimum test losses, or from a true departure of the test loss as a function of $(N,D)$ from the power law form of Equation~\ref{eq:powerlaw}.
The relatively small statistics used in our studies, the inherently large variance in minimum test loss for small models, and the fact that Equation~\ref{eq:powerlaw} has been used by Refs.~\cite{vigl2026neural} and~\cite{ATL-SOFT-PUB-2026-002} suggest that the former is more likely, but more studies would be required to confirm this hypothesis.

\begin{table}[!ht]
    \centering
    \caption{Compute-optimal scaling exponents produced by the parametric loss surface fit approach.
    Uncertainties from bootstrap resampling (100 samples, 80\% subsample fraction) of the test dataset, from which the test loss is recalculated.
    The scaling exponents are extracted from each bootstrap and the standard deviation across the ensemble is taken as the uncertainty.}
    \label{tab:exponents_approach2}
    \renewcommand{\arraystretch}{1.3}
    \begin{tabular}{lcc}
        \toprule
        Pretraining subset & $a$ ($N^*$) & $b$ ($D^*$) \\
        \midrule
        Scratch              & $0.539 \pm 0.001$ & $0.461 \pm 0.001$ \\
        QCD                  & $0.425 \pm 0.002$ & $0.575 \pm 0.002$ \\
        QCD + res2p          & $0.288 \pm 0.009$ & $0.712 \pm 0.009$ \\
        QCD + res34p         & $0.311 \pm 0.005$ & $0.689 \pm 0.005$ \\
        QCD + res2p + res34p & $0.249 \pm 0.005$ & $0.751 \pm 0.005$ \\
        \bottomrule
    \end{tabular}
\end{table}

\bibliographystyle{unsrtnat}
\bibliography{references}

\end{document}